\documentclass[12pt]{article} 

\usepackage{graphicx}
\usepackage{multicol}
\usepackage{epsfig,latexsym}

\def\lapprox{\mathrel{\mathop  {\hbox{\lower0.5ex\hbox{$\sim$}
\kern-1.1em\lower-0.7ex\hbox{$<$}}}}}
\def\gapprox{\mathrel{\mathop  {\hbox{\lower0.5ex\hbox{$\sim$}
\kern-1.1em\lower-0.7ex\hbox{$>$}}}}}

\begin{document}

\title{\Large Method to extract the primary cosmic ray spectrum 
from  very high energy $\gamma$-ray data
and its application to SNR RX J1713.7-3946}

\author{
F. L. Villante$^{1}$,  F. Vissani$^{2}$\\
$^1${\small\em Universit\`a di Ferrara and INFN, Ferrara, Italy}\\
$^2${\small\em  INFN, Laboratori Nazionali del Gran Sasso, Assergi (AQ), Italy}
}

\date{}

\maketitle

\def\abstractname{ \bf Abstract}
\begin{abstract}
{\footnotesize
Supernova remnants are likely to be the accelerators
of the galactic cosmic rays. 
Assuming the correctness of this hypothesis,  
we develop a method to extract the parent 
cosmic ray spectrum from the VHE gamma ray flux
emitted by supernova remnants (and other gamma transparent sources).
Namely, we calculate 
semi-analytically 
the (inverse) 
operator which relates an arbitrary gamma ray flux to the parent cosmic 
ray spectrum, without relying on any theoretical assumption
about the shape of the cosmic ray and/or photon spectrum.
We illustrate the use of this technique by applying 
it to the young SNR RX J1713.7-3946
which has been observed by H.E.S.S.\ experiment during 
the last three years.
Specific implementations of the method permit to use as 
an input either the parameterized VHE gamma ray flux 
or directly the raw data.  The possibility to 
detect features in the cosmic rays spectrum and 
the error in the determination of the parent cosmic ray
 spectrum are also discussed.
}
\end{abstract}



{\footnotesize
\def\contentsname{\centerline{{\small\bf  Contents}}}
\sf \tableofcontents}



\section{\sf Introduction}

There is no doubt that the main part of cosmic rays (CR) till the knee
is produced in the Milky Way~\cite{berez},  
and it seems fair to say that the favored site for CR production are the young 
supernova remnants (SNR). In fact, 
the turbulent gas of SNR is a large reservoir 
of kinetic energy~\cite{ginz}  
and this environment can support
diffusive shock waves acceleration \cite{fermi}.
In recent times, great progresses have been made
both in the observation and in the understanding of SNR.
In particular, the new generation imaging Cherenkov telescopes, in particular
H.E.S.S.~\cite{hessWEB}, 
allowed to observe the very 
high energy (VHE) gamma rays emitted by SNR, which are possibly   
generated by the decay of $\pi^0$ and $\eta$ 
produced by collision between the 
accelerated hadrons and the surrounding gas.
It is not yet possible, however, to exclude that (part of) the 
observed radiation is produced by electromagnetic processes.
In order to reach a definitive proof of the 
hadronic origin of the VHE gamma radiation,
more detailed studies are needed.
In this respect, new data at high (100 TeV or larger) and low 
($E_\gamma \sim m_\pi/2$)  
energies,
improved theoretical modeling 
and possibly observations of VHE neutrinos
(see {\em e.g.}, \cite{neutr}) will be extremely important.

The hypothesis that VHE gamma radiation from young SNR 
originates from hadronic processes
deserves the most serious attention and consideration. 
New and crucial observations are being collected and
the hadronic origin seems to be favored  
for certain SNR, such as Vela Jr \cite{vela}
and RX J1713.7-3946 \cite{rxjhadr}.\footnote{
See also \cite{waxman} for a recent analysis leading to different conclusions.}
In this paper  
we take the hadronic origin has a working hypothesis
and we address the question of what we learn on SNR cosmic ray spectra 
from VHE $\gamma-$ray data. This question has a precise quantitative character
and we answer it in the most direct way. Namely, we calculate 
semi-analytically 
the (inverse) 
operator which relates an arbitrary gamma ray flux to the parent cosmic 
ray spectrum, without relying on any theoretical assumption
about the shape of the cosmic ray and/or photon spectrum.
We then illustrate the possible applications of our method
by considering the H.E.S.S.\ data of RX J1713.7-3947 that reached 
an impressive 
accuracy in the energy range from 300 GeV 
to 300 TeV \cite{rxj}.
We remark that in this case (and, more in general, whenever the source shows non 
trivial spectral features) the approximation of power law distribution and 
the many techniques of calculations tailored to this assumption are not 
adequate.

The plan of the paper is as follows.
In Sect.~\ref{sec:2} we formulate the problem
and we obtain a general, analytical solution. 
In Sect.~\ref{sec:3} we consider possible applications 
of our results. First, we derive the parent cosmic ray flux 
of RX J1713.7-3946 by using suitable parameterizations 
of the gamma ray data. Then, we extract the information directly
from the observational data. This second technique requires fewer assumptions
and allows to propagate the observational errors 
easily. However, when applied to noisy data it requires a sort 
of image processing (Gaussian smearing) to produce a reasonable result.
In Sect.~\ref{sec:4} we summarize 
our results, putting emphasis on the 
possible applications of our method.

\section{\sf How to invert the relation between the 
photon and the CR spectrum\label{sec:2}}

\subsection{\sf Formulation of the problem}

We assume that the VHE photon flux in SNR 
has a hadronic origin, {\em i.e.}, gamma-ray are produced by a flux 
of high energy cosmic ray protons interacting with an hydrogen ambient cloud having density 
$n$. Inelastic proton-proton interactions 
result in the production of $\pi^0$ and $\eta$-mesons 
which subsequently decay producing gamma-rays. 
It is important to note that 
SNR are `transparent targets' for cosmic rays, 
as can be understood by very simple estimates. The column density 
of the system is indeed much smaller than the
TeV-proton and photon interaction lengths 
($\lambda_p\equiv m_p/\sigma \sim 40$ gr/cm$^2$ 
and $X_0\sim 60$ gr/cm$^2$), being:
$$
\begin{array}{l}
dz \equiv   n\ dl\ m_p=1.5\times 10^{-3}$ gr/cm$^2 \\
\ \  \mbox{for } n=100\mbox{ prot./cm}^3,\ dl=3\mbox{ pc}.
\end{array}
$$
The possibly overestimated value for $n$ corresponds to 
proton number density in a typical molecular cloud 
that could be associated to the SNR, 
and the distance $dl$ is the one covered in $1,000$~yr 
at a speed of $3,000$ km/s. In other words, 
the proton and photon  
interaction probabilities 
(equal to $dz/\lambda_p$ and $dz/X_{0}$ respectively) 
are 10$^{-5}$ or smaller, 
so that proton multiple interactions and/or re-absorption of the produced 
photons are absent for the typical conditions of a young SNR.

The gamma-ray flux $\Phi_{\gamma}[E_\gamma]$ produced 
on a detector placed at a distance $R$ by cosmic ray protons interacting
with a `transparent' medium can be written as\footnote{
This relation is valid if the CR momentum distribution
is approximatively isotropic. 
If this assumption is removed 
one has to replace, here and in the following:
$$
\frac{1}{4\pi}\frac{dn_{\rm p}[{\bf r},E_{\rm p}]}{dE_{\rm p}} 
\longrightarrow 
\frac{dn_{\rm p}[{\bf r},E_{\rm p},{\bf n}]}{d\Omega_{\rm p}\, dE_{\rm p}}
$$
where $dn_{\rm p} /d\Omega_{\rm p}\, dE_{\rm p}$ 
is the CR proton number density 
per unit energy and unit solid angle, ${\bf n}$ is the unit vector in 
the direction connecting the SNR to the detector and 
we have taken into account that the produced photons are almost 
collinear with CR protons.}:
\begin{equation}
\Phi_{\gamma}[E_{\gamma}]= 
\frac{c}{4\pi R^2}
\int d^3 r \; n[{\bf r}] \;  
\int_{E_{\gamma}}^{\infty}  dE_{\rm p} 
\,\frac{dn_{\rm p}[{\bf r},E_{\rm p}]}{dE_{\rm p}}
\,\frac{d\sigma_\gamma[E_{\rm p},E_\gamma]}{dE_\gamma}
\label{Initial} 
\end{equation}
where $d\sigma_\gamma/dE_\gamma$ 
is the inclusive cross-section for
$\gamma$ production. Here $n$ and $dn_{\rm p}/dE_{\rm p}$ 
are the target hydrogen number density and 
the cosmic ray proton number density (per unit energy) respectively.
Both of them depend on the position inside the SNR,
indicated by the coordinate vector ${\bf r}$. 

Next, we adopt  the usual definition of the adimensional 
distribution function $F_{\gamma}\left[x,E_{\rm p}\right]$, according 
to which:  
\begin{equation}
\frac{d\sigma_\gamma}{dE_\gamma}=\frac{\sigma[E_{\rm p}]}{E_{\rm p}}  \;
F_{\gamma}\left[\frac{E_\gamma}{E_{\rm p}},E_{\rm p}\right]
\label{Fdef}
\end{equation}
where $\sigma$ is the total inelastic p-p cross-section given 
by~\cite{kelner}:   
\begin{equation}
\sigma[E_{\rm p}] = 
34.3+1.88\ \ln[E_{\rm p}/1{\rm TeV}] + 0.25\ \ln[E_{\rm p}/1{\rm TeV}]^2 \;\; {\rm mb}
\end{equation}
Hadronic interactions are affected by quite large uncertainties and 
independent calculations of $F_{\gamma}\left[x,E_{\rm p}\right]$ may 
differ at the 20\% level \cite{tesiml}. 
In this work, we use a simple analytic formula 
presented in \cite{kelner} (see appendix \ref{AppA}) 
which describes the results of public available SIBYLL code~\cite{Sibyll}
with a few per cent accuracy over a large region of the 
parameter space ($x\ge10^{-3},E_{\rm p}> 100 \, {\rm GeV}$).
By using rel.~(\ref{Fdef}), we can rewrite eq.~(\ref{Initial}) as:
\begin{equation}
\Phi_{\gamma}[E_{\gamma}]= 
\int_{E_{\gamma}}^{\infty}  \frac{dE_{\rm p}}{E_{\rm p}} \;  
\Phi_{\rm p}[E_{\rm p}]\;
F_{\gamma}\left[\frac{E_{\gamma}}{E_{\rm p}},E_{\rm p}\right]
\label{direct1}
\end{equation} 
where we introduce the important quantity: 
\begin{equation}
\Phi_{\rm p}[E_{\rm p}]= \frac{c \; \sigma[E_{\rm p}]}{4\pi R^2}\int d^3 r \; n[{\bf r}] 
\,\frac{dn_{\rm p}[{\bf r},E_{\rm p}]}{dE_{\rm p}}
\label{PhiP} 
\end{equation}
The function $\Phi_{\rm p}[E_{\rm p}]$ 
is the quantity which is 
constrained by 
and most directly related 
to the VHE gamma ray observations. 
It has the dimensions of a differential flux, and 
below we will use $\mbox{ cm}^{-2}\mbox{ s}^{-1}\mbox{ TeV}^{-1}$.
In the following, we will refer to $\Phi_{\rm p}[E_{\rm p}]$ as the 
{\it effective cosmic ray flux from the SNR} and we will show how 
this quantity can be calculated starting from the 
photon flux $\Phi_\gamma[E_\gamma]$.
 
When comparing with theoretical predictions, one should note that
the effective CR flux encodes not only the energy distribution of cosmic ray
protons but also the (weak) energy dependence of the cosmic ray 
interaction probability in the SNR. We note, in fact, 
that eq.~(\ref{PhiP}) can be rewritten as:
\begin{equation}
\Phi_{\rm p}[E_{\rm p}]=
\frac{c\, N\,\sigma[E_{\rm p}]}{4 \pi R^2}\;J_{\rm p}[E_{\rm p}]
\label{Jp} 
\end{equation}
where $N = \int d^3 r \; n[{\bf r}]$ is the total 
amount of target hydrogen in the observed system
and $J_{\rm p}[E_{\rm p}]$ given by:
\begin{equation}
J_{\rm p}[E_{\rm p}] = \frac{1}{N}\; 
\int d^3 r \; n[{\bf r}] \;
\frac{dn_{\rm p}[{\bf r},E_{\rm p}]}{dE_{\rm p}} 
\label{Jpdef}
\end{equation}
is the weighted average the CR energy distribution {\em in} the SNR
with a weight function proportional to the target hydrogen distribution. 
A part from the constant term $c\, N / (4\pi R^2)$ (which can be deduced
if independent information on the SNR distance and on the amount of target hydrogen 
are given), the two functions $\Phi_{\rm p}[E_{\rm p}]$ and 
$J_{\rm p}[E_{\rm p}]$ differs by the energy-dependent factor $1/\sigma[E_{\rm p}]$. 
It should be noted that the cross section is slowly varying with energy, 
so that the main spectral features of $\Phi_{\rm p}[E_{\rm p}]$ 
(such as the position and the sharpness of a cutoff/transition) 
always reflects the spectral features of $J_{\rm p}[E_{\rm p}]$.
In the region where the spectrum is approximated by a power law, 
the energy dependence of $\sigma[E_{\rm p}]$ accounts for a small difference 
between the spectral indices of $J_{\rm p}[E_{\rm p}]$ and $\Phi_{\rm p}[E_{\rm p}]$ 
which can be easily quantified being of the order of 
$d\ln \sigma/d\ln E_{\rm p}\sim 0.07$ in the energy range of interest.
It is clear that the above formalism 
can be applied to any gamma transparent source (not only 
to SNR) where the VHE gamma are of hadronic origin, such 
as a SNR-molecular cloud association~\cite{gabici}.

The question of what can we learn on the effective cosmic
ray flux $\Phi_{\rm p}$ from $\Phi_\gamma$ boils down to 
the task of inverting  an {\em integral} equation (eq.~(\ref{direct1})).
In the rest of Sect.~\ref{sec:2}, we argue that, assuming a quasi-scaling
behavior of hadronic cross sections (accurate at the few percent level),  
this problem can be solved in good approximation 
 by applying the {\em differential} operator:
\begin{equation}
{\cal D}=\sum_{n=0}^5 a_n \left(E\frac{d}{dE} \right)^n
\label{diff}
\end{equation}
where $a_n$ are appropriate numerical coefficients given in sect.~\ref{sec:23}. 
Stated more clearly, we claim that the approximate inverse of eq.~(\ref{direct1}), 
symbolically written as $\Phi_\gamma={\cal F}[\Phi_{\rm p}]$, 
is simply given by $\Phi_{\rm p}\approx {\cal D}[ \Phi_\gamma]$.
This result does not rely on any theoretical 
assumption about the shape of the cosmic ray 
and/or photon spectrum. 
We thus provide a simple method 
to extract and study 
possible  spectral features of the parent cosmic ray flux in the SNR 
directly from the observed VHE gamma radiation. 

Finally, to help the readers who are more interested in applications 
than in the formal derivation of our results, 
we anticipate that, to tackle the 
mathematical problem, it was necessary 
to introduce a number of definitions.
In this paper, we indicate  the natural logarithm 
of proton and photon energies with the symbols $\varepsilon_{\rm p}$ and 
$\varepsilon_{\gamma}$ (see eq.~(\ref{energy}) in the 
next section).
Moreover, it is convenient to multiply the cosmic ray and 
the photon fluxes by a power laws in energy according to 
$\varphi_{\rm p} = \Phi_{\rm p} \cdot (E_{\rm p}/ 1 {\rm TeV})^\alpha$ and 
$\varphi_\gamma = \Phi_\gamma \cdot (E_\gamma/1 {\rm TeV})^\alpha$, 
where $\alpha$ is an appropriate coefficient. 
We remark that the 
``fluxes" $\varphi_{\rm p}$ and $\varphi_{\gamma}$ 
are related by a differential operator of the kind (\ref{diff}) for any value of $\alpha$.
The values of the numerical coefficients 
$a_n$ can be easily calculated for any adopted value for $\alpha$
(see eq.~(\ref{alpha-changes}) and related discussion).

\subsection{\sf Notation and {\em quasi scaling} approximation}

It is useful to perform some changes of variables and 
rewrite the integral (\ref{direct1}) as:
\begin{equation}
\varphi_\gamma[\varepsilon_\gamma]=  
\int_{-\infty}^\infty {d\varepsilon_{\rm p}}\  
\varphi_{\rm p}[\varepsilon_{\rm p}] \ 
f[\varepsilon_\gamma-\varepsilon_{\rm p},\varepsilon_{\rm p}]
\label{direct2}
\end{equation}
where proton and photon energies are expressed in terms of the variables:
\begin{equation}
\varepsilon_{i}=\ln\left[\frac{E_{i}}{\mbox{1 TeV}}\right], \;\;\;\;\;\; i={\rm p},\gamma
\label{energy}
\end{equation}
The fluxes are rewritten in terms of:
\begin{equation}
\left\{
\begin{array}{l}
\varphi_{\rm p}[\varepsilon_{\rm p}]
=\Phi_{\rm p}[e^{\varepsilon_{\rm p}}]\ e^{\alpha\varepsilon_{\rm p}}, \\[1ex]
\varphi_{\gamma}[\varepsilon_\gamma]
=\Phi_\gamma[e^{\varepsilon_\gamma}]\ e^{\alpha\varepsilon_\gamma},
\end{array}
\right.
\label{fluxes}
\end{equation}
and the integral kernel $f[y,\varepsilon_{\rm p}]$ is defined by:
\begin{equation}
f[y,\varepsilon_{\rm p}]=
\theta[-y] \cdot 
e^{\alpha y} \cdot  
F_\gamma[e^{y},e^{\varepsilon_{\rm p}}]
\label{kernel} 
\end{equation}
The Heaviside function is introduced in order to integrate over 
the entire real axis. The inclusion of the exponential factors in definitions
(\ref{fluxes}) and (\ref{kernel})
is, instead, motivated by the fact that the photon and CR proton spectra 
are expected to 
decrease approximatively as power laws in energy. 
For a proper choice of the parameter $\alpha$,
the functions $\varphi_i=\Phi_i \cdot (E_i/1\ {\rm TeV})^{\alpha}$
are, thus, expected to be nearly constant in the relevant energy range,
highlighting the deviations from the pure power-law behavior.
In the following, we find it convenient to set the value:
\begin{equation}
\alpha=2.5
\end{equation}
which is particularly appropriate for the analysis of the H.E.S.S.\ gamma ray 
spectrum of the RX J1713.7-3946 supernova remnant. 
We remark that the "fluxes" $\varphi_i[\varepsilon_i]$ and the 
integral kernel $f[y,\varepsilon]$ defined above
are easily tractable, since they are 
square-integrable in $\varepsilon_i$ and $y$
respectively.


\begin{figure}[t]
\par
\begin{center}
\includegraphics[width=8.5cm,angle=0]{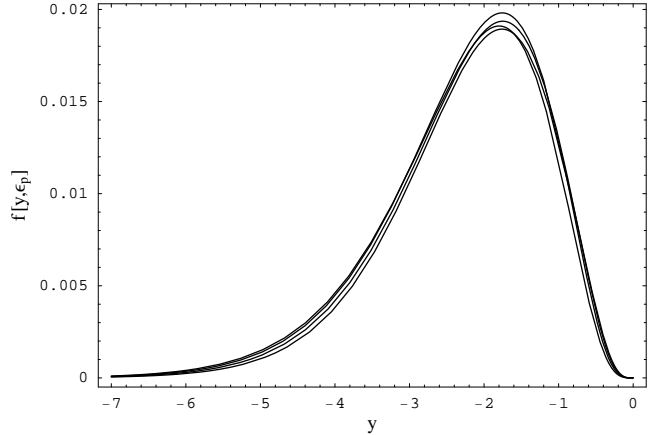}
\end{center}
\par
\vspace{-5mm} \caption{\em {\protect\small The integral kernel 
$f[y,\varepsilon_{\rm p}]$ as a function of 
$y \equiv \ln[E_\gamma/E_{\rm p}]$ for selected values of the proton energy 
$\varepsilon_{\rm p}\equiv \ln[E_{\rm p}/{\rm 1 TeV}]$. We have chosen 
$\varepsilon_{\rm p} = 0,\, 2.3,\, 4.6,\, 6.9$ 
corresponding to $E_{\rm p} \simeq 1, \, 10, \, 100,\, 1000$ TeV, respectively.}}
\label{Fig1}
\end{figure}


In Fig.~\ref{Fig1}, we show the behavior of the integral kernel $f[y,\varepsilon_{\rm p}]$ as a function of $y$ 
for selected values of $\varepsilon_{\rm p}$. We see that the function $f[y,\varepsilon_{\rm p}]$
is peaked at $y = - 1.8$ (which corresponds to $E_{\rm p}/E_\gamma = \exp[-y] \simeq 6$)
and that it is marginally dependent on the assumed proton energy. 
In the following, we assume a {\em quasi-scaling} behavior
for hadronic cross sections, {\em i.e.}, we
neglect the dependence of $f[y,\varepsilon_{\rm p}]$ on $\varepsilon_{\rm p}$ and replace:
\begin{equation}
f[y,\varepsilon_{\rm p}] \to f[y]\equiv f[y,\varepsilon_{\rm p}^{\mbox{\tiny 0}}]
\label{quasi-scaling}
\end{equation}
 where $\varepsilon_{\rm p}^{\mbox{\tiny 0}}$ is a fixed reference value for 
the proton energy. We have chosen the value $\varepsilon_{\rm p}^{\mbox{\tiny 0}}= 6.9$ 
({\em i.e.}, $E_{\rm p}^0=1000$ TeV) which is appropriate to calculate the gamma ray flux in the 
energy region $E_\gamma\simeq1-1000$ TeV probed by the
H.E.S.S.\ experiment. Our calculations and Fig.~\ref{Fig1} show that the 
quasi-scaling approximation is adequate at the few percent level.

\subsection{\sf Formal solution of the inverse problem\label{sec:23}}

\begin{figure}[t]
\par
\begin{center}
\includegraphics[width=6.5cm,angle=0]{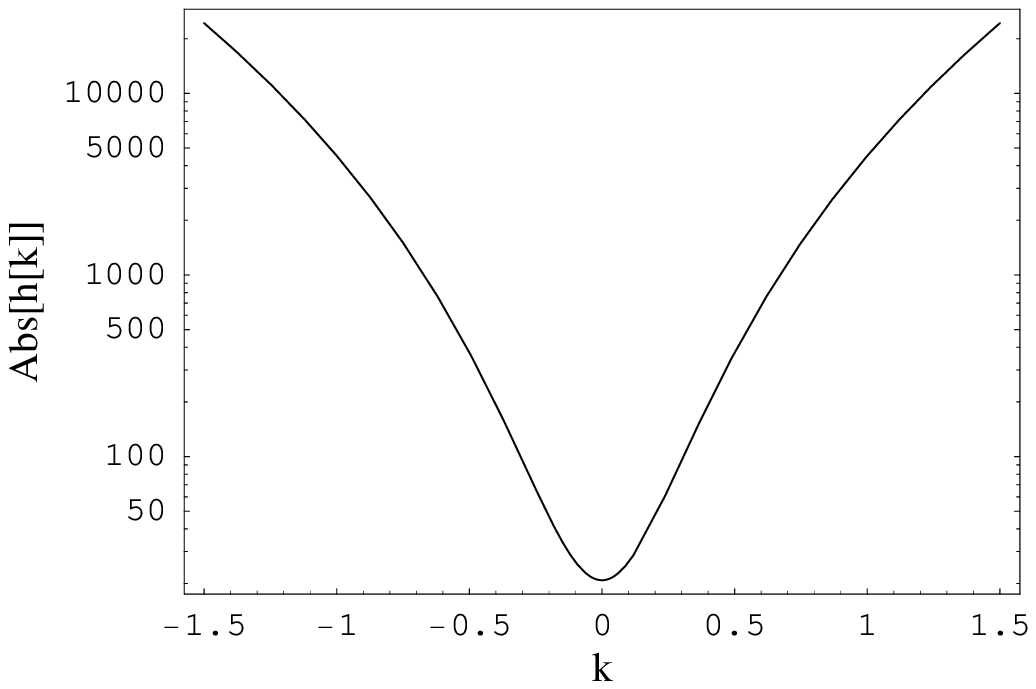}
\includegraphics[width=6.5cm,angle=0]{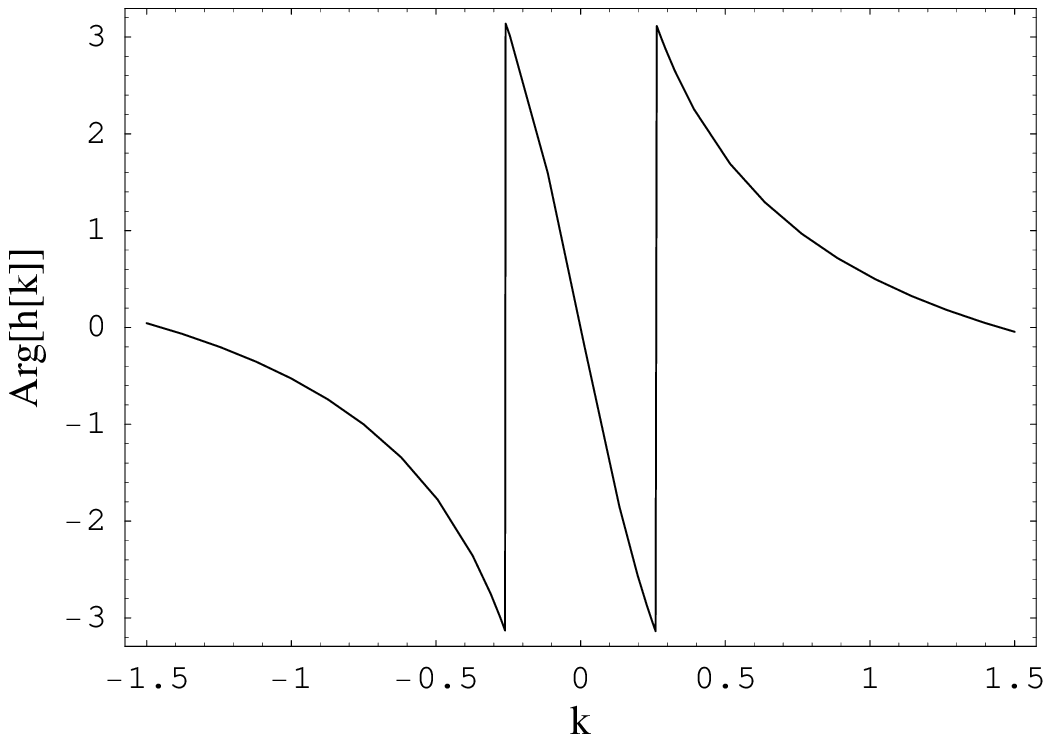}
\end{center}
\par
\vspace{-5mm} \caption{\em {\protect\small Absolute value and argument 
of the function $h[k]$ defined in eq.~(\ref{h-def}).}}
\label{Fig2}
\end{figure}

In the quasi-scaling approximation, we can invert 
the relation between the effective CR proton flux and
the photon flux (a Volterra integral equation of the first type) by a simple 
semi-analytical method which gives very precise results. 
We obtain, in fact, a convolution integral:
\begin{equation}
\varphi_\gamma[\varepsilon_\gamma]= 
\int_{-\infty}^\infty {d\varepsilon_{\rm p}}\  
\varphi_{\rm p}[\varepsilon_{\rm p}] \ 
f[\varepsilon_\gamma-\varepsilon_{\rm p}]
\label{convolution}
\end{equation}
which can be treated by using standard techniques, 
such as Fourier analysis,\footnote{We calculate Fourier transforms 
according to the standard definition: 
$\varphi[\varepsilon]=\int dk \, \varphi[k]\exp[2\pi i k \varepsilon]$.} 
finding
\begin{equation}
\varphi_{\rm p}[k] = \frac{1}{f[k]} \varphi_\gamma[k] 
\label{GammaProtFourier}
\end{equation}
where $f[k]$, $\varphi_\gamma[k]$ and $\varphi_{\rm p}[k]$
are the Fourier transforms of the functions $f[y]$, $\varphi_{\gamma}[\varepsilon_{\gamma}]$ and 
$\varphi_{\rm p}[\varepsilon_{\rm p}]$ 
respectively.
We note that 
the inclusion of the exponential factor in the definition of the
$\varphi_{\rm i}[\varepsilon_{i}]$ ensures that the Fourier transforms 
$\varphi_\gamma[k]$ and $\varphi_{\rm p}[k]$ exist.\footnote{
The functions $\varphi_{\rm i}[\varepsilon_{\rm i}]$ decrease exponentially 
for $|\varepsilon_{i}|\rightarrow \infty$ provided 
that the differential energy spectra $\Phi_{\rm i}[E_{\rm i}]$ decrease slower
that $E_{\rm i}^{-\alpha}$ at low energy and faster than $E_{\rm i}^{-\alpha}$ at high energy.}
The function:
\begin{equation}
h[k] \equiv \frac{1}{f[k]}
\label{h-def}
\end{equation}
defines, in Fourier space, the operator which inverts eq.~(\ref{convolution}).
We see from  Fig.~\ref{Fig2} that ${\rm Abs}[h[k]]$ is fast increasing with $|k|$. This can 
be understood in simple terms 
by noting that the integral kernel $f[y]$ has a half-width-half-maximum 
equal approximatively to 
$\delta y\simeq 1.0$ (see Fig.~\ref{Fig1}).
Correspondingly, the Fourier transform $f[k]$ 
has a characteristic width $\delta k\simeq 1 /(2\pi \delta y)\sim 0.16$
and its inverse function $h[k]$ has a sharp increase for $|k|\ge\delta k$.
In physical terms, this has the important consequence that   
any feature in the photon spectrum on scales smaller than 
$\delta \varepsilon_{\gamma} \le \delta y $ will be greatly amplified 
in the parent CR proton spectrum (more discussion in the next section). 


The behavior of $h[k]$ at large $k$ depends on the regularity of $f[y]$ and its 
derivatives. In this case, we can expand
$h[k]$ to fifth order in a Taylor series:
\begin{equation}
h[k]\simeq\sum_{j=0}^5 h_j \, k^j
\label{expansion}
\end{equation}
where $h_j = (1/j!) \, d^j h / d k ^j |_{k=0}$, 
with a few per cent accuracy in the relevant range $|k|<2.5$. 
We remark that the photon flux $\varphi_\gamma[\varepsilon_\gamma]$ is sampled 
by H.E.S.S.\ experiment in bins $\delta \varepsilon_\gamma=\delta E_\gamma/E_\gamma \simeq 0.2$ 
or larger 
and, thus, only the region
$|k| < 1/(2\delta \varepsilon_\gamma) = 2.5$ carries physical 
information.\footnote{
The physically significant range $|k| < 2.5$
has been estimated by applying the sampling theorem.}

The expansion (\ref{expansion}) allows to express the parent proton spectrum
as a function of the photon flux and its derivatives. By exploiting the 
properties of Fourier transforms one easily obtains:
\begin{equation}
\varphi_{\rm p}[\varepsilon_{\rm p}]=
\sum_{\rm j=0}^5\; a_j\ 
\frac{d^j\varphi_\gamma}{d\varepsilon_{\gamma}\,^j}[\varepsilon_\gamma=\varepsilon_{\rm p}] 
\label{p-from-gamma}
\end{equation}
where $a_j= h_j/ (2\pi i)^j$. The coefficients $a_j$ depend on the value of 
the parameter $\alpha$ adopted in eqs.~(\ref{fluxes},\ref{kernel}).
In our case ($\alpha=2.5$) the relevant coefficients are given by 
$a_0 = 20.85$, 
$a_1/a_0 = -2.336$, 
$a_2/a_0 = 2.113$, 
$a_3/a_0= -0.9034$, 
$a_4/a_0 = 0.1718$ and 
$a_5/a_0 =-9.79\cdot 10^{-3}$.
For a different choice $\alpha\to \alpha-\beta$, 
the coefficients $a_i$ have to be replaced by:
\begin{equation}
a_i \to  \sum_{j=i}^5 \; \frac{j!\ \beta^{j-i}}{i!\ (j-i)!}  \ a_j
\label{alpha-changes}
\end{equation}
For instance, if we set $\beta=\alpha$ 
which corresponds to the particular situation considered in  eq.~(\ref{diff}) 
({\em i.e.}, $\varphi_{\rm p}=\Phi_{\rm p}$ 
and $\varphi_\gamma=\Phi_\gamma$), we immediately obtain
$a_0 = .1148$, 
$a_1 = 2.390$, 
$a_2 = 5.205$, 
$a_3=  4.225$, 
$a_4 = 1.031$ and 
$a_5 =-0.2041$.

The above equations are the main results of this paper and, in the next section, 
we will discuss the possible applications to real data. 
Here, we note that rel.~(\ref{p-from-gamma}) is remarkably simple if 
the photon spectrum can be approximated by a power law, 
{\em i.e.}, $\Phi_\gamma \propto E_\gamma^{-\Gamma}$ or, equivalently, $\varphi_\gamma \propto \exp 
[( \alpha-\Gamma ) \varepsilon_\gamma]$. We obtain, in fact:
$\varphi_{\rm p}[\varepsilon_{\rm p}]=
{\cal Y}[\Gamma]\;\varphi_\gamma[\varepsilon_{\rm p}]$
with 
${\cal Y}[\Gamma]\equiv
\sum_{j=0}^5\; a_{j}\,
(\alpha-\Gamma)^j$
which shows that the ratio between the effective cosmic ray flux 
and the photon flux at a fixed energy is given by the function ${\cal Y}[\Gamma]$ which 
only depends on the photon spectral index $\Gamma$. The function 
${\cal Y}[\Gamma]$ can be compared with the spectrum 
weighted moments $Z[\Gamma]$ displayed in 
Fig.~5.5 of \cite{gaisser}. We obtain a good agreement by noting that, in the assumption  of \cite{gaisser}, 
one has $Z[\Gamma]\simeq\Gamma/(2\ {\cal Y}[\Gamma]$). 

Finally, we remind that the differential operator in the r.h.s.\ 
of eq.~(\ref{p-from-gamma}) is the inverse 
of the integral operator in eq.~(\ref{convolution}), which is obtained 
in the quasi-scaling assumption for hadronic cross-section. One can go beyond this approximation by 
using eq.~(\ref{p-from-gamma}) as zero-order solution and calculating perturbatively the effects of 
deviations from the quasi-scaling assumption. We used this approach to check that
the corrections in the parent cosmic ray spectrum obtained from~(\ref{p-from-gamma}) 
are small in comparison to the hadronic cross section uncertainties. We do not discuss 
the numerical implementation here to avoid unnecessary complications, but 
we provide the relevant details in the appendix \ref{AppB}.

\section{\sf Applications\label{sec:3}}

\subsection{\sf The young SNR RX J1713.7-3946\label{sec:31}}

\begin{figure}[t]
\par
\begin{center}
\includegraphics[width=9.5cm,angle=0]{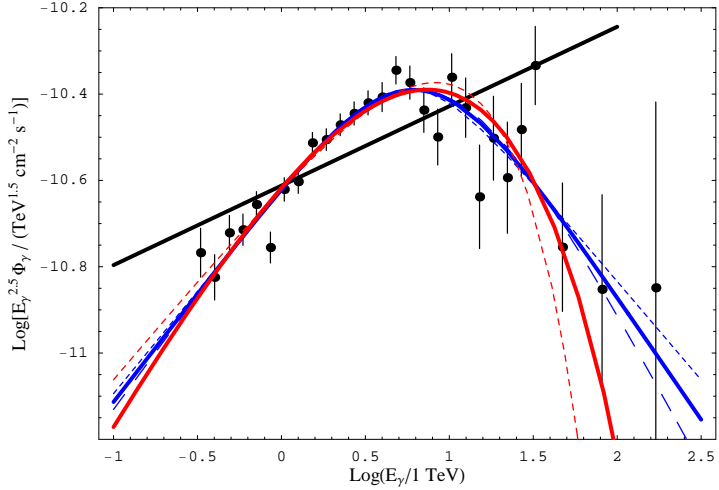}
\end{center}
\par
\vspace{-5mm} \caption{\em {\protect\small The $\gamma$-ray spectrum of the 
RX J1713.7-3946 Supernova Remnant obtained by 
H.E.S.S.\ experiment. The black line represents the best fit to the data in the assumption of a power law behavior 
of the photon spectrum. The blue lines are obtained by assuming that the photon spectrum follows the BPL given 
by eq.~(\ref{BPL}) in the assumption that the transition parameter is $S=0.6$ (blue solid line) or 
$S=0.45,\, 0.75$ (blue dotted lines). 
The red lines refers to the EC case, see eq.~(\ref{EC}), in the assumption that $\beta=0.5$ (red solid line) or $\beta=1.0$ (red dotted line).}}
\label{Fig3}
\end{figure}

The RX J1713.7-3946 Supernova Remnant has been observed by H.E.S.S.\  
during three years from 2003 to 2005 \cite{rxj}. The 
$\gamma$-ray spectrum obtained by combining the observation of the three years 
is shown in Fig.~\ref{Fig3}. 
The data extend over three decades, 
exploring the energy interval $E_\gamma = 0.3-300$ TeV.
The energy resolution of the experiment is equal to
about 20\% and the photon spectrum is sampled in 25 bins 
$\delta \varepsilon_\gamma=\delta E_\gamma/E_\gamma \simeq 0.2$ plus three larger bins at 
high energy.\footnote{To help readability, 
in Figs.~\ref{Fig3}-\ref{Fig6} we use the logarithm 
to basis 10, denoted by $\log$.}


The described data allow to obtain important conclusions,
as discussed in \cite{rxj}. First, they show 
that there is a significant 
emission at energy larger than 30 TeV, implying the existence 
of primary particles of at
least that energy. Moreover, the data show a non trivial dependence 
on energy. In particular, there is a significant deviation 
from the simple power law behavior, as can be understood from Fig.~\ref{Fig3}. 
The solid line is the best fit 
power law spectrum with spectral index $\Gamma = 2.32$ that does 
not provides an acceptable fit of the data since 
$\chi^2/{\rm d.o.f.}\sim 145.6/25$ ({\em i.e.}, can be rejected at 
$9\sigma$).   

From a theoretical point of view, one expects that the photon spectrum can be described by
power law at low energy with a ``cutoff" above a certain energy $E_{\rm c}$ 
related to to the properties of the primary particles acceleration mechanism.
 This kind of behavior is usually parameterized in the form of a broken power law (BPL):
\begin{equation}
\Phi_\gamma = I \, \left(\frac{E_\gamma}{\rm 1 TeV}\right)^{-\Gamma_1}\;  
\left(1+\left(\frac{E_\gamma}{E_{\rm c}}\right)^{1/S}\right)^{-S(\Gamma_2-\Gamma_1)}  
\;\;\;\;\;\; \mbox{ [BPL case] }
\label{BPL}
\end{equation}
where $\Gamma_1$ and $\Gamma_2$ are the low and high energy spectral indices
and $S$ quantifies the sharpness of the transition from $\Gamma_1$ and $\Gamma_2$,
or by an exponential cutoff (EC) with exponent $\beta$:
\begin{equation}
\Phi_\gamma = I \, \left(\frac{E_\gamma}{\rm 1 TeV}\right)^{-\Gamma}\; 
\exp\! \left[-\left(\frac{E_\gamma}{E_{\rm c}}\right)^\beta\right]
\;\;\; \ \ \ \ \  \ \ \ \ \   \ \ \ \  
\mbox{ [EC case] }
\label{EC}
\end{equation}
The value $\beta \sim 0.5$, which describes a relatively smooth cutoff,  
has been considered in~\cite{kelner,kappes}. 

The H.E.S.S.\ gamma ray spectrum of the RX J1713.7-3946 supernova remnant has 
been fitted with a BPL with parameters $\Gamma_1=2.00\pm0.05$, $\Gamma_2=3.1\pm0.2$ 
and $E_{\rm c}=6.6\pm2.2$ obtaining a $\chi^2/{\rm d.o.f.}\sim 29.8/23$ 
(see Fig.~\ref{Fig3}, blue solid line).
It should be noted that the sharpness parameter $S$ was kept fixed in the fit, 
with an adopted value equal to $S=0.6$. Different choices for $S$, however, 
are possible. As an example, equally good fits of 
the data are provided by the blue dashed line which corresponds to 
$S=0.75$, $\Gamma_1=1.97$, $\Gamma_2=3.22$ 
and $E_{\rm c}=7.97$ and by the blue dotted line which corresponds to 
$S=0.45$, $\Gamma_1=2.03$, $\Gamma_2=2.96$ 
and $E_{\rm c}=5.64$.

Alternatively, the H.E.S.S.\ data can be 
fitted with an EC with parameters  
$\Gamma=1.79\pm 0.06$, $E_{\rm c}=3.7\pm 1.0$ and $\beta=0.5$ 
as it is shown by the red solid line in Fig.~\ref{Fig3}, 
obtaining a $\chi^2/{\rm d.o.f.}\sim 34.3/24$. 
For comparison, we also show with a red dotted line 
the best fit which obtained by assuming ``pure" EC
({\em i.e.}, $\beta=1.0$). In this case, one obtains
 $\Gamma=2.04\pm 0.04$, $E_{\rm c}=17.9\pm 3.3$
and a slightly worse fit to the data 
$\chi^2/{\rm d.o.f.}\sim 39.5/24$
({\em i.e.}, a goodness of fit of 2.4\%).

As discussed in \cite{rxjhadr,rxj}, the observed spectral shape 
seems to favor the hadronic origin.
In the following, we assume the hadronic origin as a working hypothesis  
and we discuss what the data can tell us about the primary 
proton spectrum in RX J1713.7-3946.

\subsection{\sf Using parameterized fluxes\label{sec:32}}

\begin{figure}[t]
\par
\begin{center}
\includegraphics[width=8.5cm,angle=0]{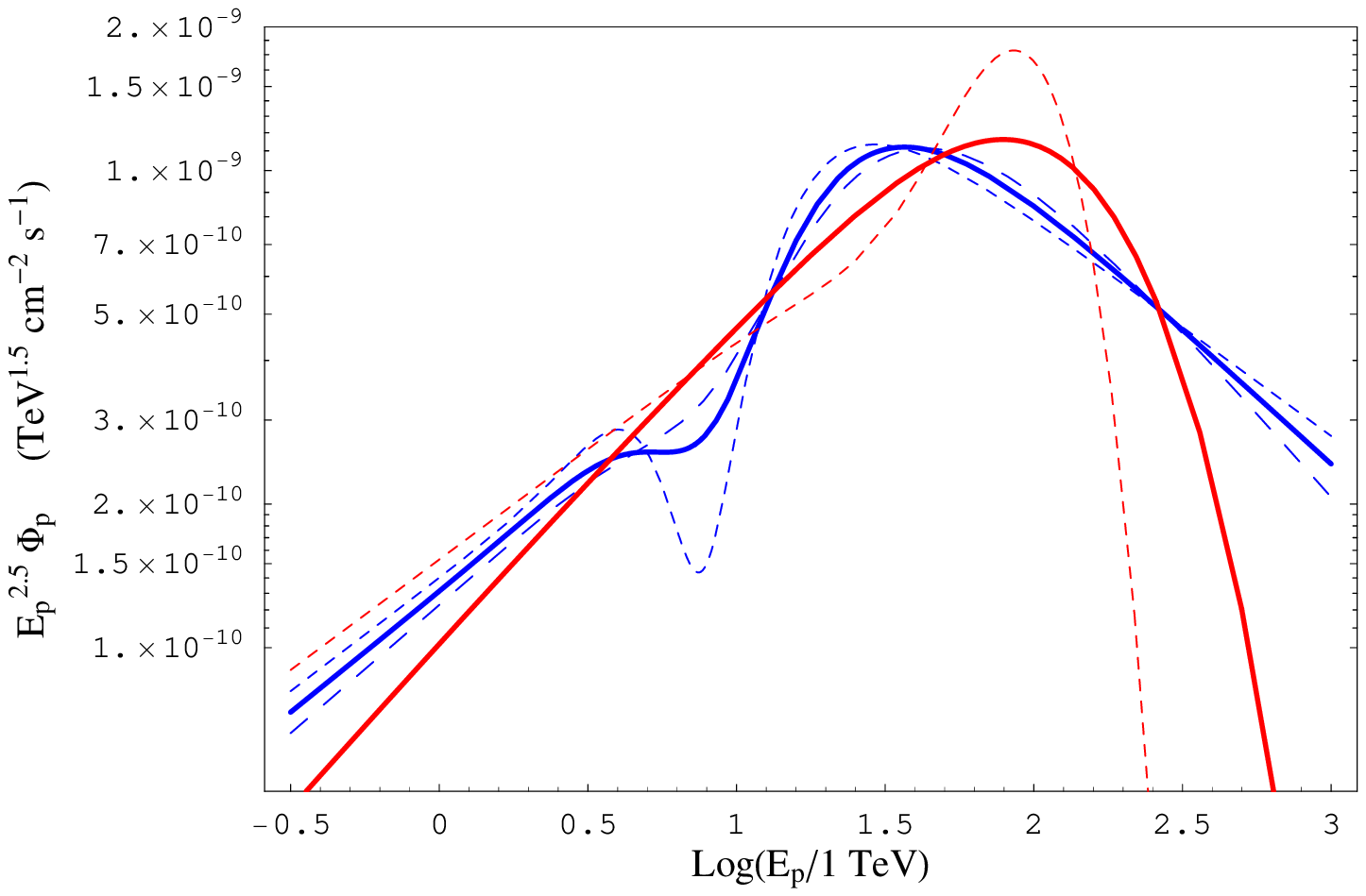}
\includegraphics[width=8.5cm,angle=0]{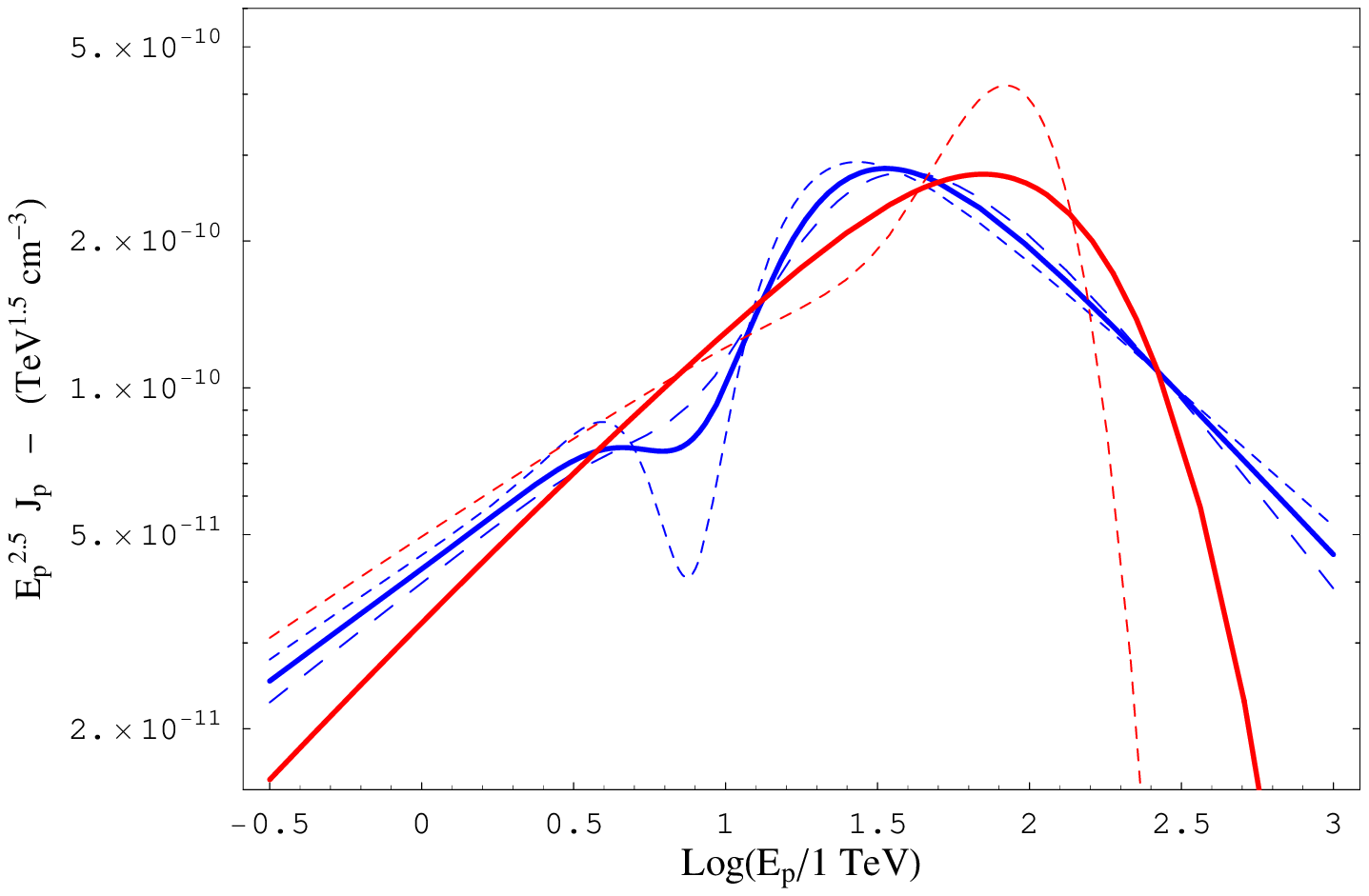}

\end{center}
\par
\vspace{-5mm} \caption{\em {\protect\small
{\sc Left Panel:} The effective CR flux $\Phi_{\rm p}[E_{\rm p}]$ from the 
SNR RX J1713.7-3946 obtained from BPL and EC parameterizations of 
the gamma-ray flux measured by the H.E.S.S.\ experiment. The blue lines are obtained 
from the best-fit BPL parameterizations with sharpness parameter $S=0.6$ (blue solid line), 
$S=0.45$ (blue dotted line) and $S=0.75$ (blue dashed line). The red lines correspond to best-fit 
EC parameterization with $\beta=0.5$ (red solid line) and $\beta=1.0$ (red dotted line).
{\sc Right Panel:} The CR energy distribution $J[E_{\rm p}]$ calculated from 
eq.~(\ref{Jp}) 
by assuming $R=1\ {\rm kpc}$ and $N=3.6\times 10^{59}$.}}
\label{Fig4}
\end{figure}

If we accept the BPL and EC parameterizations as reliable descriptions of the photon spectrum,
we can calculate the effective CR flux from the SNR by simply applying eq.~(\ref{p-from-gamma})
to the functional forms (\ref{BPL}) and (\ref{EC}).
The results of this procedure are shown in Fig.~\ref{Fig4} (left panel) 
where the blue lines are obtained from the BPL parameterizations of the photon flux, while the red lines 
refer to the EC case. We remind that the effective cosmic ray flux encodes not only the energy distribution
of cosmic ray protons but also the (weak) energy dependence of the cosmic ray
interaction probability in the SNR. For this reason we also show (right panel) 
the CR energy distribution in the SNR calculated according to eq.~(\ref{Jp}).  
We assume $R=1\ {\rm kpc}$ and $N=3.6\times 10^{59}$ which corresponds to 300 solar masses of target hydrogen. 
This value is motivated if gamma ray emission is 
due to a molecular cloud-SNR association of 
the type proposed in \cite{new1} which seems to be 
consistent with the observations of NANTEN \cite{new2};
see~\cite{th1} for a theoretical model. 
In this work, 
we do not aim to discuss the precise value of 
$N$, which would only affect 
the normalization of the CR energy distribution. 
We focus  instead on the CR spectral 
properties which are directly determined by the observed photon
spectrum. We remark a few important points.

\paragraph{\it Accuracy of the inversion.}
The obtained CR fluxes can be used in rel.~(\ref{convolution}) 
in order to check the accuracy of the inversion method. 
In all cases, the re-calculated photon fluxes agree 
with the input photon flux ({\em i.e.}, adopted in rel.~(\ref{p-from-gamma})) at the level
of few parts per thousand
in the energy range $E_\gamma=1-1000$ TeV. 
This show that the differential operator on the r.h.s.\ 
of eq.~(\ref{p-from-gamma}) is
the inverse of the integral operator in eq.~(\ref{convolution}) 
with very good accuracy, especially when compared with the uncertainties in the 
hadronic cross-section (at the $\sim 20\%$ level) or with the accuracy of the
quasi scaling approximation (at the level of few percent or better).

\paragraph{\it Cutoff/transition in the CR spectrum.}
The calculated CR spectra indicate, in all cases,
that there is a significant number of protons at high energy. 
Protons should be efficiently accelerated up to an energy equal to about 
$\sim 100$ TeV, in order to explain the observed data.
We see that the cutoff/transition region in the CR spectrum 
is in the energy region $E_{\rm p}=30-150$ TeV.
It is interesting to note that a photon flux with a smooth EC ($\beta=0.5$)  
corresponds to an effective CR flux $\Phi_{\rm p}[E_{\rm p}]$ which is well described by 
the simple functional form  $E_{\rm p}^{-\Gamma} \exp[-E/E_{\rm c}]$ 
with $\Gamma\simeq 1.79$ and cutoff energy  $E_{\rm c}\simeq 113$ TeV, in reasonable agreement 
with the conclusion of \cite{kappes}. The corresponding CR energy 
distribution $J_{\rm p}[E_{\rm p}]$ is well fitted by the same functional 
form, with the same cutoff energy and with $\Gamma\simeq1.86$.
Compare with the discussion after eq.~(\ref{Jpdef}).

\paragraph{\it Other features in CR and photon spectrum.}
We note that the differences between the CR spectra are much larger 
than the differences between the input photon fluxes. In the BPL case, 
the calculated CR spectra have a complex behavior
in the energy range $E_{\rm p}=3-30$~TeV. Similarly, in the EC case 
the obtained curves differ substantially in the energy region 
$E_{\rm p}=30-150$ TeV. 
This is not an 
artifact of the inversion method which is accurate 
at the level of few parts per thousand or better. 
It simply reflects the fact that any sharp 
feature in the photon flux is amplified in 
the parent CR spectrum. In particular, 
the sharper is 
the transition/cutoff in the photon spectrum 
({\em i.e.}, the smaller is the $S$ in eq.~(\ref{BPL}) or the larger is the 
$\beta$ in eq.~(\ref{EC})), the more complex is the behavior of the calculated
CR flux.

\paragraph{\it Dilution of spectral features.}
The previous point can be understood 
in terms of the properties of hadronic cross sections. 
It is basically related to the fact that the photon 
spectrum {\em is not supposed to have 
any sharp feature if it is originated by hadronic processes}. 
The integral kernel $f[y]$ has, in fact, a 
characteristic width $\delta y \sim 1.0$.
Consequently, features in the CR spectrum 
on scales $\delta \varepsilon_{\rm p} \le \delta y$
are washed-out  by convolution (\ref{convolution}). 
Conversely, if we observe features in the photon spectrum on scales 
$\delta \varepsilon_\gamma \le \delta y$
we are forced by rel.~(\ref{p-from-gamma}) 
to postulate a complicated behavior of the 
parent CR proton flux which may be difficult or impossible to justify.
As an example, BPL fits of the observational data with $S\le 0.4$ correspond 
to parent CR spectra which become negative in the region $E_{\rm p}=3-30$ TeV
and are, thus, not acceptable.

\paragraph{\it A plausibility test for the hadronic origin assumption.}
The presence (or the absence) of features in the observed photon flux on scales 
$\delta \varepsilon_\gamma \le \delta y$ may be used, 
in principle, as plausibility criterion to reject (or support) the hadronic origin 
of the observed $\gamma$-ray fluxes.\footnote{
The energy resolution of the H.E.S.S.\ 
experiments ($\delta \varepsilon_\gamma\simeq 0.2$)
is sufficiently good, in principle, to test whether 
there is some sharp feature in the photon spectrum.} 
Interestingly, in the case of EC parameterization, the experimental data 
prefers a smooth cutoff ($\beta\sim 0.5$) which
corresponds to a simple behavior of the primary photon flux. 
In the BPL case, our results shows instead that statistical 
errors are too large to arrive at any relevant conclusion about the sharpness of the transition. 
Thus, the fine structures of the primary CR spectra in the energy range $E_{\rm p}=3-30$ TeV
are not significantly constrained by the data.

\subsection{\sf Using the raw data}

The differential operator on the r.h.s.\ 
of eq.~(\ref{p-from-gamma}) is the inverse of the integral operator 
in eq.~(\ref{convolution}) and it allows
to obtain the CR spectrum directly from the photon flux, 
independently of any theoretical assumption.
The CR spectrum, however, depends on high
order derivatives of the photon flux which are generally known with
bad accuracy and, moreover, introduce complicated correlations
between the values of the CR flux extracted at two different energies.
This makes difficult to infer the parent CR flux 
directly from noisy data and one could be tempted to conclude  
that a parameterization of the gamma ray flux is, 
in fact, necessary. 
In this section we propose a non-parametric procedure 
(based on Gaussian smearing) that avoids these difficulties
and moreover permits to evaluate the error 
on the inferred CR spectrum.

\paragraph{\it The `smoothing' procedure.}
The relevance of the high order terms in rel.~(\ref{p-from-gamma}) 
depends on the scale of the features that we probe in the CR spectrum.
If we are interested in scales $\delta \varepsilon_{\rm p} \ge \delta$, 
we can define the {\em smoothed CR spectrum} as follows:
\begin{equation}
\overline{\varphi}_{\rm p}[\varepsilon_{\rm p},\delta]=
\int_{-\infty}^{\infty} d\varepsilon \; 
 \varphi_{\rm p}[\varepsilon]\;
r\left[\varepsilon_{\rm p}-\varepsilon,\delta\right]
\label{p-averaged}
\end{equation}
where:
\begin{equation}
r[\varepsilon,\delta]=\frac{1}{\sqrt{2\pi}\delta}\exp\left[-\frac{\varepsilon^2}{2\delta^2}\right]
\end{equation}
In Fourier space, this is equivalent to apply a Gaussian 
filter to $\varphi_{\rm p}[k]$ with a width equal to 
$\Delta k = 1/(2\pi\delta)$. We remind, for clarity, 
that $\varphi_{\rm p} = \Phi_{\rm p} \ (E_{\rm p}/{\rm 1 TeV})^{2.5}$
where the effective cosmic ray flux $\Phi_{\rm p}$ is defined 
in eq.~(\ref{PhiP}). 
Equivalently, we can define the 
{\em smoothed CR energy distribution} by:
\begin{equation}
\overline{\jmath}_{\rm p}[\varepsilon_{\rm p},\delta]=
\int_{-\infty}^{\infty} d\varepsilon \; 
{\jmath}_{\rm p}[\varepsilon]\;
r\left[\varepsilon_{\rm p}-\varepsilon,\delta\right]
\label{jsmooth}
\end{equation}  
where ${\jmath}_{\rm p} = J_{\rm p} \times (E_{\rm p}/{\rm 1 TeV})^{2.5}$
and $J_{\rm p}$ is given in eq.~(\ref{Jpdef}).
By using eq.~(\ref{Jp}), it is possible to show that:
\begin{equation}
\overline{\jmath}_{\rm p}[\varepsilon_{\rm p},\delta] \simeq
\frac{4 \pi R^2}{c\, N\,\sigma[\varepsilon_{\rm p}]}\;
\overline{\varphi}_{\rm p}[\varepsilon_{\rm p},\delta]
\label{JpSmooth} 
\end{equation}
with a few per cent accuracy, in the energy range of interest. 
We will then focus on the calculation of 
$\overline{\varphi}_{\rm p}$, showing that it can be simply estimated
from observational data.

\paragraph{\it The `smoothed' CR spectrum.}
Applying the differential operator of 
eq.~(\ref{p-from-gamma}), we find that the 
smoothed CR spectrum $\overline{\varphi}_{\rm p}$ 
is related to the photon flux by a convolution integral:
\begin{equation}
\overline{\varphi}_{\rm p}[\varepsilon_{\rm p},\delta]=
\int_{-\infty}^{\infty} d\varepsilon_\gamma \; 
\varphi_{\gamma}[\varepsilon_\gamma]\;
\rho\left[\varepsilon_{\rm p}-\varepsilon_\gamma,\delta\right]
\label{smoothed}
\end{equation}
where the convolving function $\rho[\varepsilon,\delta]$ is given by:
\begin{equation}
\rho[\varepsilon,\delta]=r[\varepsilon,\delta]
\ \sum_{i=0}^5 A_{i}\,\varepsilon^i 
\end{equation}
and the coefficients $A_{i}$ are equal to:
\begin{equation}
\begin{array}{rcl}
\displaystyle
A_0 = a_0 - \frac{a_2}{\delta^2} + 3\frac{a_4}{\delta^4}, &
\displaystyle
A_1 = - \frac{a_1}{\delta^2} + 3\frac{a_3}{\delta^4}-15\frac{a_5}{\delta^6}, &
\displaystyle
A_2 =  \frac{a_2}{\delta^4} - 6\frac{a_4}{\delta^6},\\[2ex] 
\displaystyle
A_3 =  - \frac{a_3}{\delta^6} + 10\frac{a_5}{\delta^8}, &
\displaystyle
A_4 = \frac{a_4}{\delta^8}, &
\displaystyle
A_5 = -\frac{a_5}{\delta^{10}}.
\end{array}
\end{equation}
We can apply rel.~(\ref{smoothed}) directly to the raw data, as 
it is explained in the following. We indicate 
with $\varphi_i\pm\Delta\varphi_i$ 
the value of the photon flux measured in the $i$-th bin, centered at the 
photon energy $\varepsilon_i$ and covering the energy range
$(\varepsilon_{i,{\rm inf}},\varepsilon_{i,{\rm sup}})$. We
approximate the photon flux by:
\begin{equation}
\varphi_\gamma[\varepsilon_\gamma]=\sum_i \varphi_i \, W_i[\varepsilon_\gamma]
\label{rawdata}
\end{equation}
where $W_i[\varepsilon_\gamma]$ are rectangular functions describing the various energy bins ({\em i.e.}, 
$W_i[\varepsilon_\gamma]\equiv1$ for $\varepsilon_{i,\rm inf}\le\varepsilon_\gamma\le\varepsilon_{i,\rm sup}$ 
and zero elsewhere).
We immediately obtain 
from eq.~(\ref{smoothed}) the relation:
\begin{equation}
\overline{\varphi}_{\rm p}[\varepsilon_{\rm p},\delta]=
\sum_i \varphi_i \; w_i[\varepsilon_{\rm p},\delta]
\label{fromrawdata} 
\end{equation}
where:
\begin{equation}
w_i[\varepsilon_{\rm p},\delta]=\int_{\varepsilon_{i,{\rm inf}}}^{\varepsilon_{i,{\rm sup}}} d\varepsilon_\gamma \; 
\rho\left[\varepsilon_{\rm p}-\varepsilon_\gamma,\delta\right]
\simeq 
\rho\left[\varepsilon_{\rm p}-\varepsilon_i,\delta\right] \Delta\varepsilon_i
\end{equation}
and, in the last step, we have assumed that $\delta\gg\Delta \varepsilon_i=\varepsilon_{i,{\rm sup}}-\varepsilon_{i,{\rm inf}}$. Eq.~(\ref{fromrawdata}) 
gives the desired expression of the (smoothed) CR flux 
direcly from the gamma ray data.

The smoothed CR spectrum is a 
linear combination of the observational 
values $\varphi_{i}$ of the photon flux.
The functions $w_i[\varepsilon_{\rm p},\delta]$ 
describe the contribution that each 
data point give to the reconstructed spectrum at a fixed energy $\varepsilon_{\rm p}$.
The uncertainty in the CR spectrum can be easily evaluated by propagating 
linearly the observational errors $\Delta \varphi_i$ obtaining: 
\begin{equation}
\frac{\Delta \overline{\varphi}_{\rm p}[\varepsilon_{\rm p},\delta]}
{\overline{\varphi}_{\rm p}[\varepsilon_{\rm p},\delta]} = 
\frac{\sqrt{\sum_i  \Delta\varphi_i^2  \,w_i[\varepsilon_{\rm p},\delta]^2}}
{\sum_i \varphi_i  \,w_i[\varepsilon_{\rm p},\delta]}
\label{error}
\end{equation} 
Similarly, the correlation between the values of the CR flux 
at two different energies can be obtained by calculating:
\begin{equation}
\varrho[\varepsilon_{\rm p},\varepsilon'_{\rm p},\delta] = 
\frac{\sum_k  \Delta\varphi_k^2  \ w_k[\varepsilon_{\rm p},\delta]\ w_k[\varepsilon'_{\rm p},\delta]}
{\sqrt{\sum_i  \Delta\varphi_i^2  \ w_i[\varepsilon^{}_{\rm p},\delta]^2}\
 \sqrt{\sum_j  \Delta\varphi_j^2  \ w_j[\varepsilon'_{\rm p},\delta]^2}} 
\label{correlation}
\end{equation}

\paragraph{\it Application to the RX J1713.7-3946 observations.}
Before applying the above relations to the H.E.S.S.\ data 
we have to choose the smoothing scale $\delta$. 
The choice of $\delta$ is somewhat arbitrary and depends on the 
detector, on the quality of the observational data and 
on the problem under consideration. 
The H.E.S.S.\ detector has an energy resolution equal to $\delta \varepsilon_\gamma=0.2$ 
that suggests to adopt $\delta\gg 0.2$.
Moreover, if we {\em accept} the hadronic origin 
assumption, we know that the hadronic interactions themselves 
introduce a scale, $\delta y\sim 1.0$, below which the photon spectrum is not expected
have features large enough to be significant with respect to observational errors.  
At the same time, we know that 
``noise" at these small scales is greatly amplified in the calculated CR spectrum. 
All this suggests to choose $\delta=\delta y = 1.0$ and to focus our attention
on the large scale features of the parent CR flux.

\begin{figure}[t]
\par
\begin{center}
\includegraphics[width=8.5cm,angle=0]{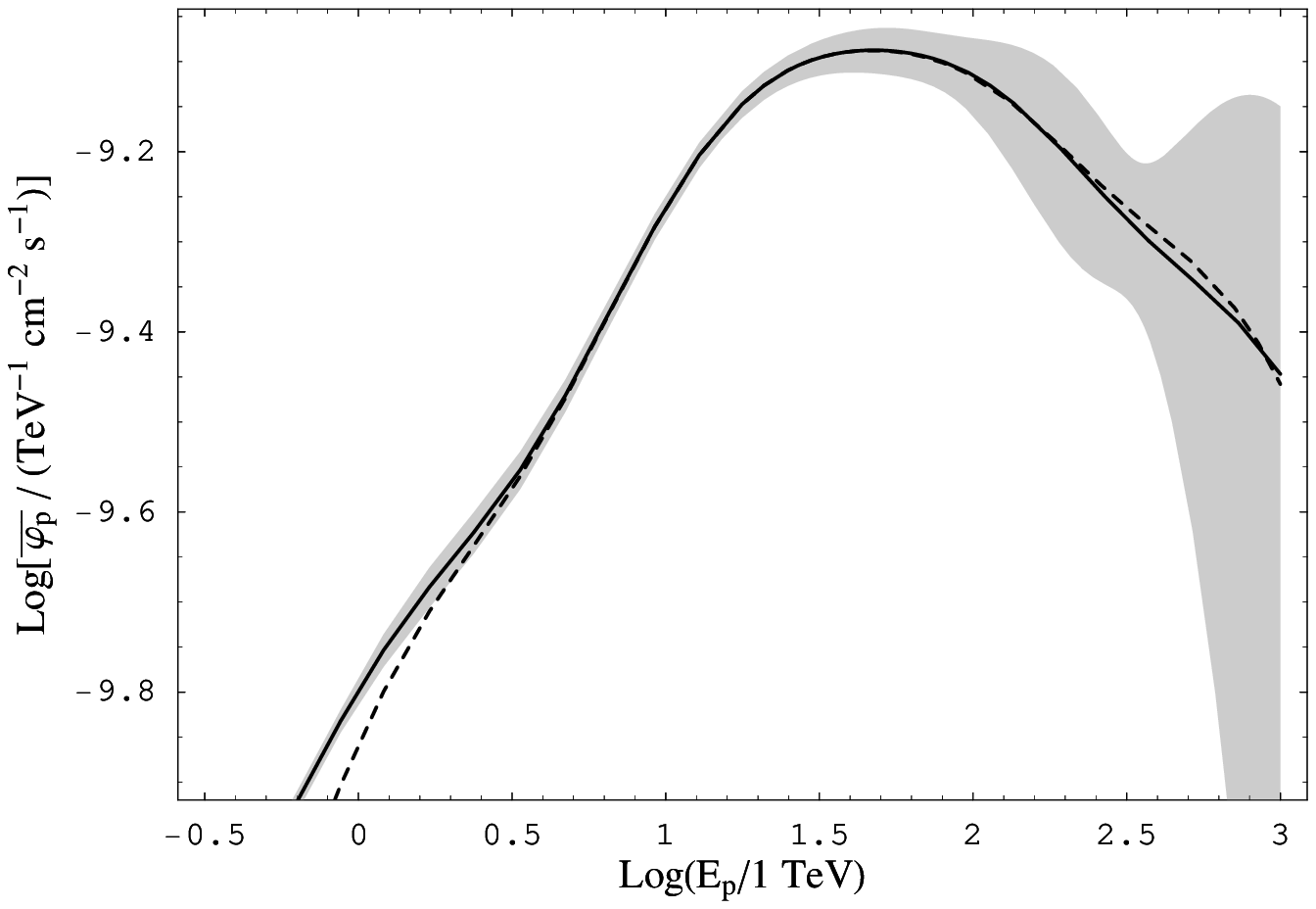}
\includegraphics[width=8.5cm,angle=0]{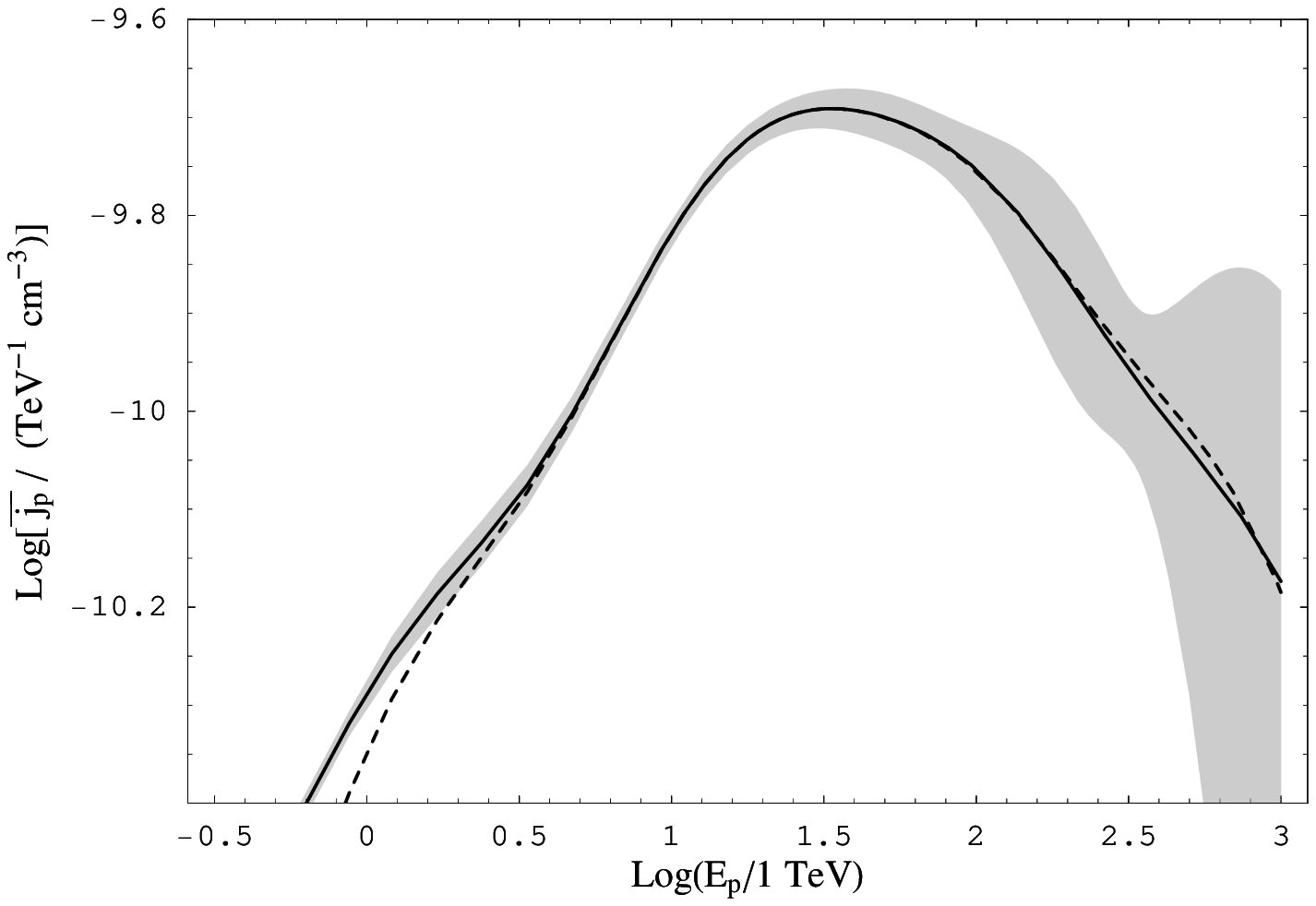}
\end{center}
\par
\vspace{-5mm} \caption{\em {\protect\small {\sc Left Panel:} The (smoothed) CR spectrum from the RX J1713.7-3946 SNR
deduced from the raw data of the H.E.S.S.\ experiments. The solid line is obtained by continuing the 
$\gamma$-ray spectrum at low and high energy  with the best-fit BPL with sharpness parameter 
$S=0.6$, while the dashed line is obtained by using the best-fit EC with $\beta=0.5$. 
The shaded area represents the observational uncertainty and is obtained by propagating H.E.S.S.\  
observational errors. {\sc Right Panel:} The (smoothed) CR energy distribution calculated from eq.~(\ref{JpSmooth})
with $R=1\ {\rm kpc}$ and $N=3.6\times 10^{59}$.}}
\label{Fig5}
\end{figure}

\begin{figure}[t]
\par
\begin{center}
\includegraphics[width=8.5cm,angle=0]{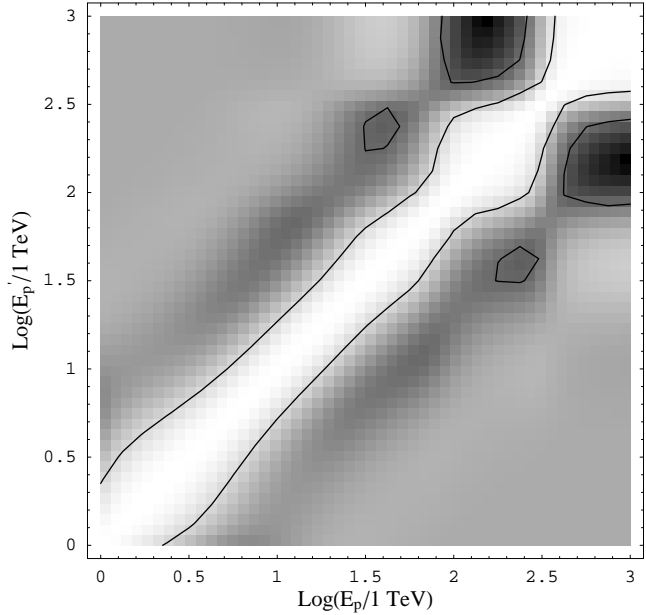}
\end{center}
\par
\vspace{-5mm} \caption{\em {\protect\small Correlation between the values of the (smoothed) proton spectrum
obtained at two different energies. Light colors correspond to values of the correlation index close to 1, 
while dark colors correspond to values close to -1. The displayed contours corresponds to correlation 
index equal to 0.5 and to -0.5.}}
\label{Fig6}
\end{figure}

Our final results are displayed in Fig.~\ref{Fig5}. In the left panel we show the smoothed CR spectrum
which is obtained from the H.E.S.S.\ observational 
data of the RX J1713.7-3946 SNR. In the right panel we show the
smoothed CR energy distribution estimated according to eq.~(\ref{JpSmooth}) 
with $R=1\ {\rm kpc}$ and $N=3.6\times 10^{59}$.
We remind that the observational 
data cover an 
energy range 
equal to $E_\gamma=0.3-300$ TeV. 
In this energy range we have described the photon spectrum according 
to eq.~(\ref{rawdata}), while at low ($E_\gamma\le 0.3$ TeV) and high energy ($E_\gamma\ge 300$ TeV) 
we have continued the photon spectrum by using 
the best-fit BPL with sharpness parameter 
$S=0.6$ (see previous section).\footnote{
Strictly speaking, one should know the photon flux at 
all energies in order use the rel.~(\ref{fromrawdata}).}
The shaded area describes 
the {\em observational uncertainty} in the smoothed CR spectrum
which is obtained by propagating the errors in the observational data according to 
eq.~(\ref{error}). One sees that the error is less than 10\% at low energy and remains smaller 
than 20\% for $E_{\rm p}\le 300$ TeV. The correlation between the values of the CR flux 
at two different energies, evaluated as in eq.~(\ref{correlation}),
is shown in Fig.~\ref{Fig6}.

In order to estimate the {\em systematic uncertainty}  
introduced by the ignorance of the high and/or low energy behaviour 
of the photon flux, we also show with a dashed line the smoothed CR spectrum
which is obtained by continuing the 
photon spectrum at high and low energy with the best-fit EC with $\beta=0.5$.
The difference between the solid and the dashed line is smaller than the
observational error for $E_{\rm p}=3-1000$ TeV, showing that the proton spectrum, 
in this energy range, is directly constrained by observational data. 
For $E_{\rm p}\le 3\ {\rm TeV}$, instead, the two lines behave differently
indicating that the systematic uncertainty due to extrapolation is relevant. 
In this respect, the small bend of the 
effective CR flux in the energy region $1-10$ TeV, at the moment, does not 
seem to be fully signicative. The existence of such a 
bend would amount to an important physical information 
on the acceleration mechanism
(see {\em e.g.}, \cite{blasi}). 
Thus, it will be interesting to collect
new data at energy lower than $E_\gamma=0.3$ TeV to assess its 
significance.

In conclusion, the displayed results show that the large scale 
features of the CR spectrum in the energy range $E_{\rm p}=3-300$ 
TeV are well constrained by the observational data.
The effective CR flux is roughly described 
by power-law with spectral index $\Gamma = 1.7-2$ at low energy 
with a cutoff/transition region between $E_{\rm p}=30-100$ TeV.
This conclusion is also consistent with the one obtained in
Sect.~\ref{sec:32} using parameterized photon fluxes.

\section{\sf Summary\label{sec:4}}
  
In this work we assumed the hadronic origin of the gamma radiation emitted 
by SNR and we addressed the question of what can be learned on the SNR
cosmic ray spectrum from VHE $\gamma$-ray data. We summarize here our 
conclusions:

\medskip
{\it i)} The main result is 
contained in eq.~(\ref{p-from-gamma}). This equation shows that, 
in the approximation of quasi-scaling behavior of the 
hadronic cross sections,\footnote{The 
``quasi scaling" assumption  
defined in eq.~(\ref{quasi-scaling}) 
is accurate at the few per cent level, see Fig.~\ref{Fig1}, 
and can be improved as discussed in the appendix \ref{AppB}.} the 
effective CR spectrum defined in eq.~(\ref{PhiP}) 
can be obtained by applying a simple differential operator
to the observed photon flux. 
This results does not rely on any 
theoretical assumption about the shape of the 
proton and/or photon spectrum.  
It can thus be 
applied to
sources which show non trivial spectral features 
such SNR RX J1713.7-3946 for which, instead, the commonly adopted 
approximation of power law distribution and 
the many techniques of calculations tailored to this case are not
adequate.

\medskip
{\it ii)}  
We have emphasized that the 
presence (or the absence) of sharp features in the photon spectrum can be 
used as plausibility criterion to reject 
(or to support) the assumption that the observed radiation has a hadronic origin. 
The basic point is that the hadronic 
processes (convolved with the parent CR flux) lead to a characteristic 
energy scale below which the produced
photon flux is expected to be featureless 
(see discussion in Sect.~\ref{sec:32}).

\medskip
{\it iii)} Specific implementations of our 
method permit to calculate the parent 
CR spectrum either from parameterized VHE fluxes or directly from raw data 
(see eq.~(\ref{fromrawdata})). This second approach requires fewer theoretical 
assumptions and allows to propagate the observational errors easily 
(see eqs.(\ref{error},\ref{correlation})).   
However, when applied to noisy data, it requires a sort of image 
processing (Gaussian smearing, see eq.~(\ref{p-averaged})) to 
produce reasonable results.

\medskip
{\it iv)} 
We have applied our method to 
the young SNR RX J1713.7-3946
which has been observed by the H.E.S.S.\ experiment 
during the last three years. 
We have calculated the CR spectrum both from 
parameterized photon fluxes and 
directly from the raw data. The results are summarized in 
Fig.~\ref{Fig4} and Fig.~\ref{Fig5}. 
These figures demonstrate that the observational data 
constrain well the main features 
of CR flux in the energy 
range $E_{\rm p}\simeq 3 - 300$ TeV; they give, instead,
a poor information outside this range, and cannot significantly 
test the fine structures of the CR spectrum. 
As a final result, 
we conclude that the effective CR flux from SNR RX J1713.7-3946
is well described by power-law with spectral index $\Gamma = 1.7-2$ 
at low energy with a cutoff/transition region between $E_{\rm p}=30-100$ TeV. 

\subsection*{\sf Acknowledgments} 

We are grateful to 
P.~Blasi, M.L.~Costantini and P.~Lipari for 
useful discussions
and to the anonymous referee for useful suggestions.
This work was supported by the High Energy Astrophysics Studies
Contract No. ASI-INAF I/088/06/0, by the MIUR grant PRIN 2006
"Astroparticle Physics" and by the European FP6 Network "UniverseNet"
MRTN-CT-2006-035863



\appendix\section{\sf The functions $F_\gamma[E_\gamma/E_{\rm p},E_{\rm p}]$ and $f[y]$\label{AppA}} 

 In order to give a self-contained discussion, we give here the function
$F_\gamma[E_\gamma/E_{\rm p},E_{\rm p}]$ obtained in~\cite{kelner} 
and used in this paper. We have:
\begin{eqnarray}\nonumber
F_\gamma[x,\,E_{\rm p}]= B_\gamma\,\frac{\ln[x]}{x}\left(\frac{1-x^{\beta_\gamma}}
{1+k_\gamma x^{\beta_\gamma}(1-x^{\beta_\gamma})}\right)^4
\left(\frac1{\ln[x]}-\frac{4\beta_\gamma x^{\beta_\gamma}}{1-x^{\beta_\gamma}}
-\frac{4k_\gamma\beta_\gamma x^{\beta_\gamma}(1-2x^{\beta_\gamma})}
{1+k_\gamma x^{\beta_\gamma}(1-x^{\beta_\gamma})}
\right)
\end{eqnarray}
where $x=E_\gamma/E_{\rm p}$.
The parameters $B_\gamma$, $\beta_\gamma$, and $k_\gamma$
depend only on the energy of proton and are given by:
\begin{eqnarray}
B_\gamma&=&1.30+0.14\,\varepsilon_{\rm p}+0.011\,\varepsilon_{\rm p}^2\,,\nonumber\\
\beta_\gamma&=&\frac1{1.79+0.11\,\varepsilon_{\rm p}+0.008\,\varepsilon_{\rm p}^2}\,\nonumber,\\
k_\gamma&=&\frac1{0.801+0.049\,\varepsilon_{\rm p}+0.014\,\varepsilon_{\rm p}^2}\,\nonumber,
\end{eqnarray}
where $\varepsilon_{\rm p}=\ln[E_{\rm p}/1\,{\rm TeV}]$. The function $f[y]$
is obtained by applying eqs. (\ref{kernel}) and (\ref{quasi-scaling}) and is simply given by:
$$
f[y]=\theta[-y]\cdot e^{\alpha y}\cdot F_\gamma[e^y,1 {\rm PeV}]
$$

\section{\sf Improving the {\em quasi scaling} 
approximation \label{AppB}}
Let us begin by writing the integral equation eq.~(\ref{PhiP})
in abstract terms:
\begin{equation}
\Phi_\gamma[E_\gamma] = {\cal F}[\Phi_{\rm p}][E_\gamma]
\end{equation}
We showed that 
a solution of the integral equation 
$\Phi_\gamma={\cal F}_0[\Phi_{\rm p}]$
in  the quasi-scaling approximation
$F[x,E_{\rm p}]\to F[x,E_{\rm p}^0]$,  
accurate at the level of few parts per thousand or better, is given by 
\begin{equation}
\Phi_{\rm p}={\cal D}[\Phi_\gamma]
\label{solutia}
\end{equation}
In operator terms, we 
write ${\cal D}{\cal F}_0={\cal F}_0{\cal D}=1$.
For a given gamma-ray flux $\Phi_\gamma$ we can evaluate  the goodness
of a certain approximation of the proton `flux' $\Phi_{\rm p}$ 
from the difference between the assumed 
gamma-ray flux and the gamma-ray flux  
re-calculated from the proton `flux':
\begin{equation}
\xi=\frac{{\cal F}[\Phi_{\rm p}]-\Phi_\gamma}{\Phi_\gamma}
\label{corria}
\end{equation}
In our case, we evaluate the goodness of the 
quasi-scaling approximation by using eq.~(\ref{solutia})
for $\Phi_{\rm p}.$
Assuming that $\Phi_\gamma$ is a broken power-law, 
we find that $\xi$ is smaller than 10\% 
in the range of energies from 100 GeV to 1 PeV,  
that is already sufficient 
for our purposes. It is however possible to improve 
the approximation for the proton 
flux as follows:
\begin{equation}
\Phi_{\rm p}={\cal D}[\Phi_\gamma (1-\xi)]
\label{improvia}
\end{equation}
where $\xi$ is calculated 
using eq.~(\ref{solutia}).\footnote{The 
formal derivation is simple: From 
$\Phi={\cal F}[\Psi]=({\cal F}_0+\delta {\cal F})[\Psi]$ we get
${\cal F}_0[\Psi]=\Phi-\delta {\cal F}[\Psi]$. Applying 
${\cal D}$ we find 
$\Psi={\cal D}[\Phi -\delta{\cal F}[\Psi]],$ 
that can be improved 
iteratively. If in the r.h.s.\ of the last equation
we use $\Psi={\cal D}[\Phi]$ ({\em i.e.}, eq.~(\ref{solutia}))
we get eq.~(\ref{improvia})
simply applying the definition of eq.~(\ref{corria}).}
Repeating the procedure 
(namely: assuming again the broken power-law distribution for $\Phi_\gamma$
and plugging eq.~(\ref{improvia}) into eq.~(\ref{corria})) in order 
to test the approximation, the newly calculated $\xi$ 
is smaller than 0.5\% in the range from 100 GeV to 1 PeV.

\newpage

 \footnotesize 

\section*{\sf  References}
\def\refname{


\end{document}